\documentclass[twocolumn,aps,prc,showpacs,10pt]{revtex4-1}

\usepackage{amsmath}
\usepackage{amssymb}
\usepackage{graphicx}
\usepackage{mathrsfs}
\begin{document}
\title{Description of single-$\Lambda$ hypernuclei with relativistic point coupling model}

\author{Y. Tanimura}
\author{K. Hagino}
\affiliation{Department of Physics, Tohoku University, Sendai 980-8578, Japan}

\date{\today}
%\maketitle
%\pagenumbering{roman}
%\tableofcontents
%\newpage
%\pagenumbering{arabic}

\begin{abstract}
We extend the relativistic point coupling model to single-$\Lambda$ hypernuclei. 
For this purpose, we add $N$-$\Lambda$ effective contact couplings to the model Lagrangian, 
and determine the parameters by fitting to the experimental data for $\Lambda$ binding energies. 
Our model well reproduces the data over a wide 
range of mass region although some of our interactions yield the reverse ordering 
of the spin-orbit partners 
from that of nucleons for heavy hypernuclei. 
The consistency of the interaction with the quark model predictions is also discussed.
\end{abstract}

\pacs{21.80.+a,23.20.Lv,21.30.Fe,21.60.Jz}

\maketitle

\begin{section}{INTRODUCTION}
Relativistic mean field (RMF) theory has been successfully 
applied to both finite nuclei and nuclear matter in order to 
describe their bulk properties \cite{SeWa,VALR05,P96,MTZZLG06}. 
Starting from an effective Lagrangian in which 
nucleon and meson fields are coupled in a covariant manner, 
single-particle Dirac equations for nucleons are derived within the 
mean field approximation.
In this model, 
nucleons are treated as Dirac particles moving independently in 
the mean field generated by the mesons.
The spin-orbit interaction with the 
correct sign and magnitude naturally 
arises from the relativistic treatment of nucleons.
The success of the model affirms the meson exchange picture 
of a nucleon-nucleon interaction in nuclei.

Recently, another class of relativistic model, that is, 
the relativistic point coupling (RPC) model \cite{MaMa,NiHoMa} 
has also been widely employed
 \cite{BuMaMaRe,MaBuMa,SuBuMaReGr,NiVrRi,YMRV10,NLVPMR09,ZhLiYaMe}. 
This model consists of Skyrme-type zero-range interactions 
and has been found to be as 
capable 
as meson 
exchange RMF models of reproducing the properties of 
finite nuclei and nuclear matter \cite{NiHoMa,BuMaMaRe,NiVrRi,YMRV10,NLVPMR09}.
This model has 
several advantages compared to the meson exchange models.
First, there is no need to solve the Klein-Gordon equations 
for mesons since the mesonic degrees of freedom are all 
implicit in the RPC model. 
Secondly, the Fock terms can easily be introduced by using the Fierz 
transformation 
because of its zero-range nature \cite{MaBuMa,SuBuMaReGr}. 
Lastly, it is much easier to apply the model to 
beyond-mean-field methods 
such as the generator 
coordinate method (GCM), and 
angular momentum and particle number projections \cite{YMRV10,NLVPMR09}.

A zero-range type interaction is suitable also for 
the three-dimensional (3D) mesh method for mean-field calculations 
in the coordinate space representation \cite{BFHKW85,ev8}.  
With this method, 
an arbitrary deformation of nuclei can be efficiently described, and 
the method has been widely used in non-relativistic 
Skyrme Hartree-Fock (SHF) calculations 
together with the imaginary time technique \cite{BFHKW85,ev8}. 
An extension of this method to the relativistic approach is not trivial, however. 
That is, a naive 
imaginary time evolution breaks down in relativistic systems due to the 
presence of the Dirac sea \cite{ZhLiMe,HagiTani}. 
This is not a numerical, but rather a fundamental problem 
related to the variational principle.
Such phenomena have been well-known under the name of ``variational collapse'' 
in the field of relativistic quantum chemistry \cite{WaKu81,WaKu83,StHa,HiKr,FaFoCh}.
For this reason, a 3D mesh calculation has not yet been carried out neither with RMF nor 
RPC. Recently, a few prescriptions to avoid the variational collapse have been tested 
in the nuclear physics context \cite{ZhLiMe,HagiTani}. 
The prescriptions were found to work well, at least for simple spherical systems, 
and a 3D mesh calculation may now be almost ready to perform. 

In this paper, we extend the relativistic point coupling model to hypernuclei. 
Effects of adding a $\Lambda$ particle on the shape of nuclei, 
{\it i.e.} the glue-like role, are attracting 
much attention theoretically 
and experimentally \cite{HiKaMiMo,MyHa08,MyHaKo,BiEnSh,IsKiDoOh,Ta,YLHWZM11}.
The RMF theory with meson exchange model has been extended 
to hypernuclei in order to describe such effects by adding  
ordinary scalar and 
vector couplings of a $\Lambda$ particle to $\sigma$ and $\omega$ mesons, 
respectively 
\cite{BrWe,BoBo,Bou,MaJe,SuTo,SoYaLuMe,CoWe,Jeetal,FiKaVrWe09,MyHa08,TsMaMaOh,BiEnSh,V98}.
The authors of Refs. 
\cite{SuTo,SoYaLuMe} fitted the coupling constants 
in the strange sector to the experimental data of $\Lambda$
binding energies.
In these calculations, 
the tensor coupling to the $\omega$ meson, which is 
predicted by the quark model \cite{CoWe,Jeetal} to be much 
stronger for $\Lambda$ than for nucleon, was also included 
in order to reproduce rather small $\Lambda$ spin-orbit splittings. 
Our aim in this paper is to propose a zero-range version of those phenomenological RMF model 
for hypernuclei. 

The paper is organized as follows. In Sec. \ref{sec:model}, 
we introduce the model Lagrangian for a point coupling model 
extended to single-$\Lambda$ hypernuclei. 
In Sec. \ref{sec:result}, the optimal parameter set 
is obtained by fitting to experimental single-particle energies for the $\Lambda$ particle. 
We also test 
its predictive power and a consistency 
with the quark model. 
In Sec. \ref{sec:summary}, we summarize the paper.
\label{sec:intro}
\end{section}

\begin{section}{\label{sec:model}the MODEL}

Our model Lagrangian for single-$\Lambda$ hypernuclei is given by
\begin{equation}
{\cal L}=
{\cal L}_{\rm free}+{\cal L}_{\rm em}+{\cal L}_{\rm int}^{N\! N}+{\cal L}_{\rm int}^{N\! \Lambda},
\label{totL}
\end{equation}
in which 
the free and electromagnetic parts are given by 
\begin{eqnarray}
{\cal L}_{\rm free}&=&
\bar{\psi}_{N}(i\partial\llap /-m_N)\psi_N
+\bar{\psi}_{\Lambda}(i\partial\llap /-m_{\Lambda})\psi_{\Lambda},\\
{\cal L}_{\rm em}&=&
-e\bar{\psi}_{N}A\llap /\frac{1-\tau_3}{2}\psi_N-\frac{1}{4}F_{\mu\nu}F^{\mu\nu},
\end{eqnarray}
respectively. 
Here, $\psi_N$, $\psi_{\Lambda}$, and $A^\mu$ are the nucleon, lambda, 
and electromagnetic fields, respectively. 
$F^{\mu\nu}=\partial^{\mu}A^{\nu}-\partial^{\nu}A^{\mu}$ 
is the electromagnetic field strength tensor. 
The masses of nucleon and lambda are denoted by $m_N$ and $m_{\Lambda}$, respectively. 
$\tau_3$ is the isospin matrix. 

The RPC model for normal nuclei consists of four-fermion point 
couplings ${\cal L}_{\rm 4f}$, 
derivative terms ${\cal L}_{\rm der}$, and higher order terms ${\cal L}_{\rm hot}$ \cite{BuMaMaRe}. 
Here, ${\cal L}_{\rm 4f}$ is the leading-order of zero-range approximation to 
the meson exchange interaction. 
${\cal L}_{\rm der}$ simulates the finite ranges of the meson exchanges.
${\cal L}_{\rm hot}$ corresponds to the self couplings of the scalar 
and vector mesons, which introduces a density dependence into $N$-$N$ 
contact couplings. 
These terms are given by 
\begin{equation}
{\cal L}_{\rm int}^{N\! N}=
{\cal L}_{\rm 4f}^{N\! N}+{\cal L}_{\rm der}^{N\! N}+{\cal L}_{\rm hot}^{N\! N},
\end{equation}
with
\begin{eqnarray}
\begin{aligned}
{\cal L}_{\rm 4f}^{N\! N}=
&
-\frac{1}{2}\alpha_S(\bar{\psi}_{N}\psi_N)(\bar{\psi}_{N}\psi_N)\\
&
-\frac{1}{2}\alpha_V(\bar{\psi}_{N}\gamma_{\mu}\psi_N)(\bar{\psi}_{N}\gamma^{\mu}\psi_N)\\
&
-\frac{1}{2}\alpha_{TS}(\bar{\psi}_{N}\vec\tau\psi_N)\cdot(\bar{\psi}_{N}\vec\tau\psi_N)\\
&
-\frac{1}{2}\alpha_{TV}(\bar{\psi}_{N}\gamma_{\mu}\vec\tau\psi_N)\cdot(\bar{\psi}_{N}\gamma^{\mu}\vec\tau\psi_N),
\end{aligned}
\label{eq:NN4f}
\end{eqnarray}
\begin{eqnarray}
\begin{aligned}
{\cal L}_{\rm der}^{N\! N}=
&
-\frac{1}{2}\delta_S(\partial_{\mu}\bar{\psi}_{N}\psi_N)(\partial^{\mu}\bar{\psi}_{N}\psi_N)\\
&
-\frac{1}{2}\delta_V(\partial_{\mu}\bar{\psi}_{N}\gamma_{\nu}\psi_N)
(\partial^{\mu}\bar{\psi}_{N}\gamma^{\nu}\psi_N)\\
&
-\frac{1}{2}\delta_{TS}(\partial_{\mu}\bar{\psi}_{N}\vec\tau\psi_N)\cdot(\partial^{\mu}\bar{\psi}_{N}\vec\tau\psi_N)\\
&
-\frac{1}{2}\delta_{TV}(\partial_{\mu}\bar{\psi}_{N}\gamma_{\nu}\vec\tau\psi_N)
\cdot(\partial^{\mu}\bar{\psi}_{N}\gamma^{\nu}\vec\tau\psi_N),
\end{aligned}
\label{eq:NNder}
\end{eqnarray}
and
\begin{eqnarray}
\begin{aligned}
{\cal L}_{\rm hot}^{N\! N}=&
-\frac{1}{3}\beta_S(\bar{\psi}_{N}\psi_N)^3
-\frac{1}{4}\gamma_S(\bar{\psi}_{N}\psi_N)^4\\
&
-\frac{1}{4}\gamma_V\left[(\bar{\psi}_{N}\gamma_{\mu}\psi_N)
(\bar{\psi}_{N}\gamma^{\mu}\psi_N)\right]^2.
\end{aligned}
\label{eq:NNhot}
\end{eqnarray}
Notice that the four different spin-isospin vertex structures labeled by 
the subscripts $S$, $V$, $TS$, and $TV$ 
in the coupling constants
correspond to $\sigma$, $\omega$, $\delta$, and $\rho$ meson exchanges, respectively.
Thus we can find one-to-one correspondence of each term to the 
meson exchange model.

Noticing that $\Lambda$ only couples to the scalar and vector mesons, 
we construct $N$-$\Lambda$ interaction as
\begin{eqnarray}
\begin{aligned}
{\cal L}_{\rm int}^{N\! \Lambda}=
{\cal L}_{\rm 4f}^{N\! \Lambda}+{\cal L}_{\rm der}^{N\!\Lambda}
+{\cal L}_{\rm ten}^{N\!\Lambda},
\end{aligned}
%\label{eq:}
\end{eqnarray}
where 
\begin{eqnarray}
\begin{aligned}
{\cal L}_{\rm 4f}^{N\!\Lambda}=
&
-\alpha_S^{(N\!\Lambda)}(\bar{\psi}_{N}\psi_N)(\bar{\psi}_{\Lambda}\psi_{\Lambda})\\
&
-\alpha_V^{(N\!\Lambda)}(\bar{\psi}_{N}\gamma_{\mu}\psi_N)
(\bar{\psi}_{\Lambda}\gamma^{\mu}\psi_{\Lambda}),
\end{aligned}
\label{eq:NL4f}
\end{eqnarray}
\begin{eqnarray}
\begin{aligned}
{\cal L}_{\rm der}^{N\!\Lambda}=
&
-\delta_S^{(N\!\Lambda)}(\partial_{\mu}\bar{\psi}_{N}\psi_N)
(\partial^{\mu}\bar{\psi}_{\Lambda}\psi_{\Lambda})\\
&
-\delta_V^{(N\!\Lambda)}(\partial_{\mu}\bar{\psi}_{N}\gamma_{\nu}\psi_N)
(\partial^{\mu}\bar{\psi}_{\Lambda}\gamma^{\nu}\psi_{\Lambda}),
\end{aligned}
\label{eq:NLder}
\end{eqnarray}
and
\begin{eqnarray}
\begin{aligned}
{\cal L}_{\rm ten}^{N\!\Lambda}=
\alpha^{(N\!\Lambda)}_T(\bar{\psi}_{\Lambda}\sigma^{\mu\nu}\psi_{\Lambda})
(\partial_{\nu}\bar{\psi}_{N}\gamma_{\mu}\psi_N).
\end{aligned}
\label{eq:NLten}
\end{eqnarray}
For simplicity, we do not consider the higher order term for the 
$N\Lambda$ coupling, ${\cal L}_{\rm hot}^{N\!\Lambda}$, in this paper. 
${\cal L}_{\rm ten}^{N\!\Lambda}$ in Eq. (\ref{eq:NLten}) 
simulates the $\Lambda$-$\omega$ tensor 
coupling ${\cal L}_{\rm ten}^{\Lambda\omega}$ $=$ $\frac{f_{\Lambda\omega}}{2m_{\Lambda}}
(\bar{\psi}_{\Lambda}\sigma^{\mu\nu}\psi_{\Lambda})(\partial_{\nu}\omega_{\mu}).$
As we mentioned in the Introduction, the quark model suggests that the 
tensor coupling of $\Lambda$ to $\omega$ meson  
is much stronger than that of nucleon. 
That is, the quark model yields 
the ratio of $\Lambda$-$\omega$ tensor-to-vector coupling constants,  
$f_{\Lambda\omega}/g_{\Lambda\omega}$, to be $-1$, 
while it yields the corresponding ratio for nucleon to be 
$f_{N\omega}/g_{N\omega}=-0.09$ \cite{CoWe}. 
Thus this type of coupling plays an important role 
in hypernuclei.
Since this term is proportional to the derivative of the mean field, 
it mainly affects the spin-orbit splittings of $\Lambda$ single-particle energies \cite{SuTo}.
It is expected that the small spin-orbit splittings of lambda
 can be reproduced by tuning the tensor coupling
 $\alpha^{(N\!\Lambda)}_T$.
We will discuss this point in the next Section.

Our model presented in this paper 
is similar to the one adopted in Ref. \cite{FiKaVrWe09} by 
Finelli {\it et al.}, which is based on the chiral SU(3) dynamics.
The $N$-$\Lambda$ part of their model consists of the density 
dependent contact four fermion couplings of the scalar and the vector 
types, as well as the derivative term of the scalar type. 
A part of the four fermion terms effectively describe unresolved
 short distance physics, 
while 
the density dependence is attributed to in-medium Nambu-Goldstone 
boson (two-pion and kaon) exchanges. 
The coefficients for the density-dependence 
are fixed by the chiral SU(3) perturbation theory. 
The tensor interaction caused by $2\pi$ exchange is treated in the model of Ref. \cite{FiKaVrWe09} 
as 
first-order perturbation on the Hartree single-particle energies. 
In contrast to the model of Finelli {\it et al.}, 
we determine all the parameters phenomenologically, and thus 
we consider in our model also the derivative coupling 
of the vector type, which is absent in theirs. 

%%%%%%%%%%%%%%%%%%%%%%%%%%%%%%%%%%%%%%%%%%%%%%%%%%%%%%%%%%%%

The total energy corresponding to the Lagrangian in Eq. (\ref{totL})
for a single-$\Lambda$ hypernucleus with mass number
 $A$ ({\it i.e.,} a single $\Lambda$ particle with 
$A-1$ nucleons) 
in the mean field (Hartree) and the no-sea approximations is given by
\begin{widetext}
\begin{equation}
\begin{array}{rcl}
E&=&
\displaystyle \int d^3r\ \biggl(\sum_{i=1}^{A-1}\psi_i^{\dagger}(\vec\alpha\cdot\vec p+m_N\beta)\psi_i
+\psi_{\Lambda}^{\dagger}(\vec\alpha\cdot\vec p+m_{\Lambda}\beta)\psi_{\Lambda}
+\frac{1}{2}eA^0\rho_V^{\rm (p)}\biggr.
+\frac{1}{2}\sum_{K}\alpha_K\rho^2_K+\frac{1}{2}\sum_{K}\delta_K\rho_K\Delta\rho_K\\
&&
\displaystyle 
+\frac{1}{3}\beta_S\rho_S^3+\frac{1}{4}\gamma_S\rho_S^4+\frac{1}{4}\gamma_V\rho_V^4
\displaystyle \biggl. +\sum_{K=S,V}\alpha_K^{(N\!\Lambda)}\rho_K\rho_K^{(\Lambda)}
+\sum_{K=S,V}\delta_K^{(N\!\Lambda)}\rho_K\Delta\rho^{(\Lambda)}_K
+\alpha^{(N\!\Lambda)}_T\rho_T^{(\Lambda)}\rho_V\biggr),
\end{array}
\label{eq:Etot}
\end{equation}
\end{widetext}
where $\vec{\alpha}$ and $\beta$ are the usual Dirac matrices.
Here, we have assumed the time reversal invariance of the nuclear ground state.
The densities appearing in Eq. (\ref{eq:Etot}) are defined as
\begin{eqnarray}
\rho_S&=&\sum_{i=1}^{A-1}\bar{\psi}_i\psi_i,
~~~~~~~\rho_V=\sum_{i=1}^{A-1}\psi_i^{\dagger}\psi_i, \\
\rho_{TS}&=&\sum_{i=1}^{A-1}\bar{\psi}_i\tau_3\psi_i,
~~~\rho_{TV}=\sum_{i=1}^{A-1}\psi_i^{\dagger}\tau_3\psi_i, \\
\rho^{(\Lambda)}_S&=&
\bar{\psi}_{\Lambda}\psi_{\Lambda},~~~~~~~~~~
\rho^{(\Lambda)}_V=
\psi^\dagger_{\Lambda}\psi_{\Lambda},\\
\rho^{(\Lambda)}_T&=&\vec\nabla\cdot(\bar{\psi}_{\Lambda}i\vec\alpha\psi_{\Lambda}).
\end{eqnarray}
Here $\psi_i$ is the wave function for the $i$-th nucleon, and  
$\psi_{\Lambda}$ is the wave function for the $\Lambda$ particle.

The relativistic Hartree equations for the nucleons and lambda particle
are obtained by taking the variation of the energy with respect to the wave functions 
as,
\begin{equation}
\frac{\delta}{\delta\psi_i^{\dagger}(\vec r)}
\biggl(E-\sum_{j=1}^{A}\epsilon_j\int d^3r'\psi_j^{\dagger}\psi_j
\biggr)=0,
\end{equation}
where $\epsilon_i$ is a Lagrange multiplier which ensures 
the normalization of the single particle wave functions. 
Variation with respect to the nucleon wave function leads to 
the Hartree equation for nucleons,
\begin{equation}
\begin{array}{r}
\left[\vec\alpha\cdot\vec p+V_V+V_{TV}\tau_3+V_C
+\left(m_N+V_S+V_{TS}\tau_3\right)\beta\right]\psi_i\\
=\epsilon_i\psi_i,
\end{array}
\end{equation}
with
\begin{equation}
\begin{array}{rcl}
V_S&=&
\alpha_S\rho_S+\beta_S\rho_S^2+\gamma_S\rho_S^3+\delta_S\Delta\rho_S\\
&&
+\alpha_S^{(N\!\Lambda)}\rho_S^{(\Lambda)}+\delta_S^{(N\!\Lambda)}\Delta\rho_S^{(\Lambda)},\\
V_V&=&\alpha_V\rho_V+\gamma_V\rho_V^3+\delta_V\Delta\rho_V\\
&&+\alpha_V^{(N\!\Lambda)}\rho_V^{(\Lambda)}+\delta_V^{(N\!\Lambda)}\Delta\rho_V^{(\Lambda)}
+\alpha^{(N\!\Lambda)}_T\rho_T^{(\Lambda)},\\
V_{TS}&=&\alpha_{TS}\rho_{TS}+\delta_{TS}\Delta\rho_{TS},\\
V_{TV}&=&\alpha_{TV}\rho_{TV}+\delta_{TV}\Delta\rho_{TV},\\
V_C&=&eA^0\frac{1-\tau_3}{2},\ (\Delta A^0=-e\rho_V^{(p)}),
\end{array}
\end{equation}
while variation with respect to the lambda wave function leads to 
the Hartree equation for the lambda particle:
\begin{equation}
\left[\vec\alpha\cdot\vec p+U_V+U_T
+\left(m_{\Lambda}+U_S\right)\beta\right]\psi_{\Lambda}=\epsilon_{\Lambda}\psi_{\Lambda},
\end{equation}
with
\begin{equation}
\begin{array}{rcl}
U_S&=&\alpha_S^{(N\!\Lambda)}\rho_S+\delta_S^{(N\!\Lambda)}\Delta\rho_S,\\
U_V&=&\alpha_V^{(N\!\Lambda)}\rho_V+\delta_V^{(N\!\Lambda)}\Delta\rho_V,\\
U_T&=&-i\alpha^{(N\!\Lambda)}_T \beta\vec\alpha\cdot(\vec\nabla\rho_V).
\end{array}
\end{equation}
After having solved these Hartree equations self-consistently, 
we obtain 
the total binding energy as
\begin{widetext}
\begin{equation}
\begin{array}{rcl}
E_B&=&
\displaystyle \sum_{i=1}^{A-1}\epsilon_i+\epsilon_{\Lambda}-E_{\rm CM}-(A-1)m_N-m_{\Lambda}\\
&&
\displaystyle
-\int d^3r\ \biggl(
\frac{1}{2}\sum_{K}\alpha_K\rho^2_K+\frac{1}{2}\sum_{K}\delta_K\rho_K\Delta\rho_K
+\frac{2}{3}\beta_S\rho_S^3+\frac{3}{4}\gamma_S\rho_S^4+\frac{3}{4}\gamma_V\rho_V^4
\biggr.\\
&&
\displaystyle
\biggl.
+\sum_{K=S,V}\alpha_K^{(N\!\Lambda)}\rho_K\rho_K^{(\Lambda)}
+\sum_{K=S,V}\delta_K^{(N\!\Lambda)}\rho_K\Delta\rho^{(\Lambda)}_K
+\alpha^{(N\!\Lambda)}_T\rho_T^{(\Lambda)}\rho_V
+\frac{1}{2}eA^0\rho^{(p)}_V
\biggr),
\end{array}
\end{equation}
\end{widetext}
where the center of mass energy $E_{\rm CM}$ is calculated 
by taking the expectation value of 
the kinetic energy for the center of mass motion 
with respect to the many-body ground state wave function as 
\begin{equation}
E_{\rm CM}=\frac{\langle P_{\rm CM}^2 \rangle}{2[(A-1)m_N+m_{\Lambda}]}.
\label{cmenergy}
\end{equation}
See Appendix for the explicit expression for this term. 

The relation of the point coupling model
to the meson exchange model can be made as follows 
(See also Eqs. (6)-(10) in Ref.\cite{BuMaMaRe}). 
By eliminating the meson fields and expanding the meson 
propagators to the leading order, the following approximate 
relations between the two models can be obtained \cite{BuMaMaRe}:
\begin{equation}
\alpha_S\approx-\frac{g^2_{N\sigma}}{m_{\sigma}^2},
\ \alpha_V\approx\frac{g^2_{N\omega}}{m_{\omega}^2},
\label{eq:alpha-g}
\end{equation}
\begin{equation}
\alpha^{(N\!\Lambda)}_S\approx
-\frac{g_{N\sigma}g_{\Lambda\sigma}}{m_{\sigma}^2},
\ \alpha^{(N\!\Lambda)}_V\approx
\frac{g_{N\omega}g_{\Lambda\omega}}{m_{\omega}^2},
\ \alpha^{(N\!\Lambda)}_T\approx
-\frac{g_{N\omega}f_{\Lambda\omega}}{2m_{\Lambda}m_{\omega}^2},
\end{equation}
where $g$'s and $m$'s are the baryon-meson coupling constants 
and the meson masses, respectively. $f_{\Lambda\omega}$ is the 
$\Lambda$-$\omega$ tensor coupling constant. 
Notice that it has been demonstrated that $\alpha_S$ and $\alpha_V$ 
obtained phenomenologically approximately follow these relations 
\cite{BuMaMaRe}. 
If we assume the naive quark counting ratios
 $g_{\Lambda\sigma}=\frac{2}{3}g_{N\sigma}$ and 
$g_{\Lambda\omega}=\frac{2}{3}g_{N\omega}$, together with  
the quark model prediction for the tensor coupling, 
$f_{\Lambda\omega}/g_{\Lambda\omega}=-1$, 
we obtain 
\begin{equation}
\alpha^{(N\!\Lambda)}_S\approx\frac{2}{3}\alpha_S,\ 
\alpha^{(N\!\Lambda)}_V\approx\frac{2}{3}\alpha_V,\ 
\alpha^{(N\!\Lambda)}_T\approx-\frac{\alpha_V}{3m_{\Lambda}}.
\label{eq:2/3}
\end{equation}
We will show in the next Section that these expected relations indeed 
hold if 
we include the $N$-$\Lambda$ tensor coupling given by Eq. (\ref{eq:NLten}) 
in the Lagrangian. 

%%%%%%%%%%%%%%%%%%%%%%%%%%%%%%%%%%%%%%%%%%%%%%%%%%%%%%%%%%%
\end{section}

\begin{section}{RESULTS AND DISCUSSION}
With the model described in the previous section, 
we calculate $\Lambda$ binding energies defined by the mass difference
\begin{equation}
m(^{A-1}Z)+m_{\Lambda}-m(^A_{\Lambda}Z)=
E_{B}(^{A-1}Z)-E_{B}(^A_{\Lambda}Z). 
\end{equation}
To this end, we assume spherical symmetry, and neglect the pairing 
correlations for simplicity. 
For the valence orbit, we use the filling approximation to determine 
the occupation probability. 
We set the masses of baryons to $m_N=938\ {\rm MeV}$ and $m_{\Lambda}=1115.6\ {\rm MeV}$.
We use the parameter set PC-F1 \cite{BuMaMaRe} for the $N$-$N$ part of interaction and 
fit the five parameters in the $N$-$\Lambda$ part (see Eqs. (\ref{eq:NL4f}), 
(\ref{eq:NLder}) and (\ref{eq:NLten})) 
to the experimental data. 
The data to be fitted to are $\Lambda$ binding energies 
for $s$ and $p$ orbitals in $^{16}_{\ \Lambda}{\rm O}$, 
$s$, $p$ and $d$ in $^{40}_{\ \Lambda}{\rm Ca}$, 
$s$ and $d$ in $^{51}_{\ \Lambda}{\rm V}$, 
$s,\ p,\ d$, and $f$ in $^{89}_{\ \Lambda}{\rm Y}$, 
$s,\ p,\ d,\ f$, and $g$ in $^{139}_{\ \ \Lambda}{\rm La}$, and
$s,\ p,\ d,\ f$, and $g$ in $^{208}_{\ \ \Lambda}{\rm Pb}$. These are 
taken from Refs.\cite{HaTa,UsBo}. 
In addition, the spin-orbit splitting for the 
$p$ orbital of $\Lambda$ in $^{16}_{\ \Lambda}{\rm O}$ \cite{Mo} is 
included in the fitting procedure. 
The value deduced in Ref.\cite{Mo} 
is $300\ {\rm keV}\leq \epsilon_{\Lambda p_{1/2}}
-\epsilon_{\Lambda p_{3/2}}\leq 600\ {\rm keV}$, 
where the variation comes from a choice of the interactions. 
Notice that this value is model dependent, and we merely 
regard it as a criterion.
The coupling constants in the strange sector are determined by performing 
a least-squares fit to the data, that is, by minimizing the quantity
\begin{equation}
\chi^2_{\rm dof}=\frac{1}{N_{\rm dof}}\sum_{i=1}^N
\left(\frac{O^{\rm theor}_i-O^{\rm expt}_i}{\Delta O^{\rm expt}_i}\right)^2.
\label{eq:chi2}
\end{equation}
Here, 
$N_{\rm dof}$ is the number of degree of freedom, and $O^{\rm theor}_i$ and 
$O^{\rm expt}_i$ are theoretical and experimental values of 
the observables, respectively, with the experimental 
uncertainties of $\Delta O^{\rm expt}_i$.
To find the minimum of $\chi^2_{\rm dof}$ in the five dimensional parameter space, 
we employ an automatic search algorithm 
\textit{Oak-ridge and Oxford method} \cite{AlFeRo}. 

The parameter set PCY-S1 so obtained is summarized in Table \ref{tb:PCY-S1}.
Together with the coupling constants, the ratios $R$ of 
the resultant $N$-$\Lambda$ coupling constants to the expected 
values given in Eq. (\ref{eq:2/3}),
\begin{equation}
R=({\rm resulted\ value})/({\rm expected\ value}),
\label{eq:ratio}
\end{equation}
are also shown.
These ratios are $R=0.79$, $0.96$, and $1.37$ 
for $\alpha^{(N\!\Lambda)}_S$, 
$\alpha^{(N\!\Lambda)}_V$, and $\alpha^{(N\!\Lambda)}_T$, respectively, 
and the expected values are approximately realized.

The calculated binding energies 
of $\Lambda$ 
with this interaction are shown in the upper panel of Fig. \ref{fig:PCY-S1}. 
One observes that the calculated $\Lambda$ binding energies agree 
with the experimental values fairly well, although 
the binding energies for 
$^{28}_{\ \Lambda}{\rm Si}$ and $^{32}_{\ \Lambda}{\rm S}$ are somewhat overestimated. 
The less satisfactory result for these latter nuclei, 
which has been observed also in the previous RMF calculations for 
hypernuclei \cite{MaJe,SuTo,SoYaLuMe,TsMaMaOh}, 
is within expectation, as we 
do not take into account 
a strong deformation of the core nucleus nor the pairing correlation. 
We have confirmed that the situation does not change 
even if we include these two nuclei in the fitting. 

In order to investigate the role of the tensor coupling, 
we show in Table \ref{tb:PCY-S2} the parameter set PCY-S2 obtained \textit{without} 
including the tensor coupling term. 
The Lambda binding energies calculated with this interaction is shown in 
the upper panel of Fig. \ref{fig:PCY-S2}. As one sees, the agreement with 
the experimental data is worsened as compared to PCY-S1, 
and the ratios $R$ are strongly suppressed compared to unity. 
On the other hand, the sum $\alpha_S^{(N\!\Lambda)}+\alpha_V^{(N\!\Lambda)}$ has 
similar values around $-3\times 10^{-5}\ {\rm MeV}^{-2}$ for PCY-S1 and 
PCY-S2. 
The suppression of the ratios can be understood as follows. 
In the non-relativistic reduction of a Dirac equation without the tensor 
coupling contribution, 
the central potential and the spin-orbit potential 
read 
\begin{equation}
V_{\rm central}=V+S,\  
V_{\rm ls}=\frac{1}{2m^2}\frac{1}{r}\frac{d}{dr}(V-S),
\end{equation}
where $V$ and $S$ are the vector and the scalar potentials, respectively. 
Therefore, to reproduce a small spin-orbit splitting of $\Lambda$ without 
the tensor interaction, 
the difference of the vector and the scalar potential have to be
small, keeping their sum constant. 
This can be achieved only by lowering the values of 
the four fermion $N$-$\Lambda$ couplings, 
$\alpha_S^{(N\!\Lambda)}$ and $\alpha_V^{(N\!\Lambda)}$, which roughly 
determine the strengths of mean potential felt by $\Lambda$. 
Notice that $V-S$ does not have to be small in the presence of the 
tensor coupling, as there is another contribution to the spin-orbit potential 
from the tensor coupling. 
The importance of the $N$-$\Lambda$ 
tensor coupling (originated from the $\Lambda$-$\omega$ tensor coupling) 
is thus evident. 
It yields small spin-orbit splittings, 
keeping $\alpha_S^{(N\!\Lambda)}$ 
and $\alpha_V^{(N\!\Lambda)}$ at the natural values. 
In PCY-S1, 
the two quark model predictions, that is, the quark counting ratios and 
the importance of the tensor coupling ($f_{\Lambda\omega}/g_{\Lambda\omega}=-1$), 
are simultaneously satisfied. 

\begin{table}
\caption{The best fit parameter set PCY-S1 
for the relativistic point coupling model for hypernuclei. 
PC-F1 \cite{BuMaMaRe} is used for the $N$-$N$ part. 
The ratios $R$ defined in Eq. (\ref{eq:ratio}) with the expected values 
given in Eq. (\ref{eq:2/3}) 
are also shown in the table. 
The chi-square value per degree of freedom is 
$\chi^2_{\rm dof}=0.54$. }
\begin{center}
\begin{tabular}{ccc}
\hline\hline
coupling const. & value & $R$\\
\hline
$\alpha^{(N\! \Lambda)}_{S}$ & $-2.0305\times 10^{-4}\ {\rm MeV}^{-2}$ & 0.79\\
$\alpha^{(N\! \Lambda)}_{V}$ & $ 1.6548\times 10^{-4}\ {\rm MeV}^{-2}$ & 0.96\\
$\delta^{(N\! \Lambda)}_{S}$ & $ 2.2929\times 10^{-9}\ {\rm MeV}^{-4}$ & --\\
$\delta^{(N\! \Lambda)}_{V}$ & $-2.3872\times 10^{-9}\ {\rm MeV}^{-4}$ & --\\
$\alpha^{(N\! \Lambda)}_{T}$ & $-1.0603\times 10^{-7}\ {\rm MeV}^{-3}$ & 1.37\\
\hline\hline
\end{tabular}
\end{center}
\label{tb:PCY-S1}
\end{table}

\begin{table}
\caption{The parameter set PCY-S2  
obtained by the omitting the tensor coupling. $\chi^2_{\rm dof}$ is 0.85.}
\begin{center}
\begin{tabular}{ccc}
\hline\hline
coupling const. & value & $R$\\
\hline
$\alpha^{(N\! \Lambda)}_{S}$ & $-4.1595\times 10^{-5}\ {\rm MeV}^{-2}$ & 0.16\\
$\alpha^{(N\! \Lambda)}_{V}$ & $ 1.3402\times 10^{-5}\ {\rm MeV}^{-2}$ & 0.08\\
$\delta^{(N\! \Lambda)}_{S}$ & $ 1.3167\times 10^{-9}\ {\rm MeV}^{-4}$ & --\\
$\delta^{(N\! \Lambda)}_{V}$ & $-1.4018\times 10^{-9}\ {\rm MeV}^{-4}$ & --\\
$\alpha^{(N\! \Lambda)}_{T}$ & 0                                       & --\\
\hline\hline
\end{tabular}
\end{center}
\label{tb:PCY-S2}
\end{table}

\begin{table}
\caption{The parameter set PCY-S3 obtained 
without fitting to the spin-orbit splitting in $^{16}_{ \Lambda}{\rm O}$. 
$\chi^2_{\rm dof}$ is 0.57.}
\begin{center}
\begin{tabular}{ccc}
\hline\hline
coupling const. & value & $R$\\
\hline
$\alpha^{(N\! \Lambda)}_{S}$ & $-2.0197\times 10^{-4}\ {\rm MeV}^{-2}$ & 0.79\\
$\alpha^{(N\! \Lambda)}_{V}$ & $ 1.6449\times 10^{-4}\ {\rm MeV}^{-2}$ & 0.95\\
$\delta^{(N\! \Lambda)}_{S}$ & $ 2.3514\times 10^{-9}\ {\rm MeV}^{-4}$ & --\\
$\delta^{(N\! \Lambda)}_{V}$ & $-2.4993\times 10^{-9}\ {\rm MeV}^{-4}$ & --\\
$\alpha^{(N\! \Lambda)}_{T}$ & $-4.0820\times 10^{-9}\ {\rm MeV}^{-3}$ & 0.05\\
\hline\hline
\end{tabular}
\end{center}
\label{tb:PCY-S3}
\end{table}

\begin{figure}
\begin{center}
\includegraphics[scale=.36,angle=-90]{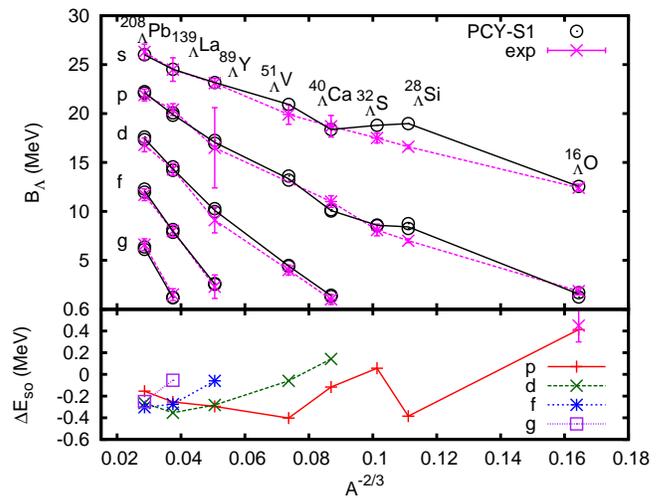}
\end{center}
\caption{(Color online) 
Comparison between 
the experimental data and 
the calculated $\Lambda$ 
binding energies $B_{\Lambda}$ (upper panel) 
and spin-orbit splittings of $\Lambda$ single-particle 
energies $\Delta E_{\rm so}$ (lower panel) obtained with 
the parameter set PCY-S1. 
The experimental data are taken from Refs.\cite{HaTa,Mo,UsBo}.}
\label{fig:PCY-S1}
\end{figure}

\begin{figure}
\begin{center}
\includegraphics[scale=.36,angle=-90]{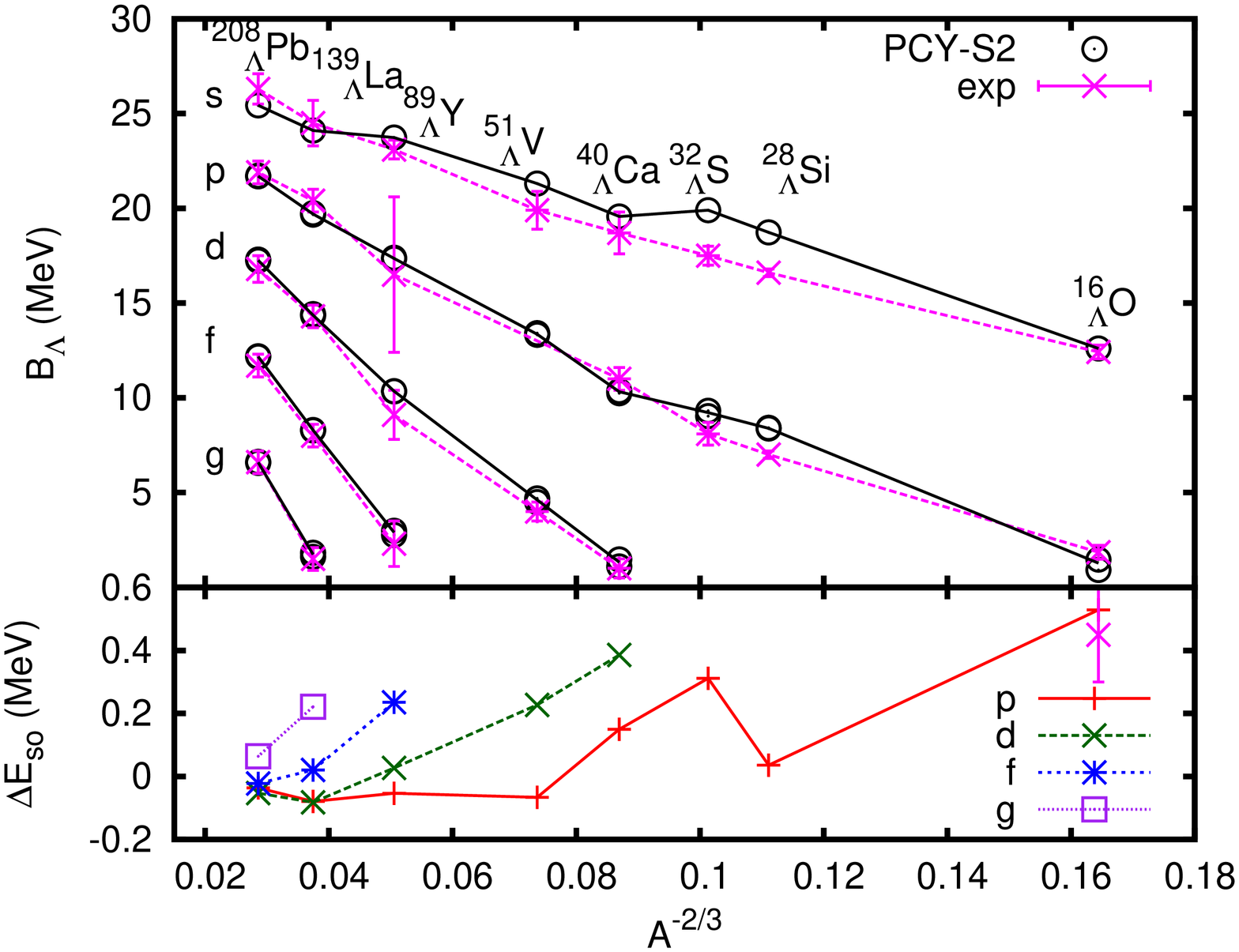}
\end{center}
\caption{(Color online) Same as Fig. \ref{fig:PCY-S1}, but with 
the parameter set PCY-S2.}
\label{fig:PCY-S2}
\end{figure}

\begin{figure}
\begin{center}
\includegraphics[scale=.36,angle=-90]{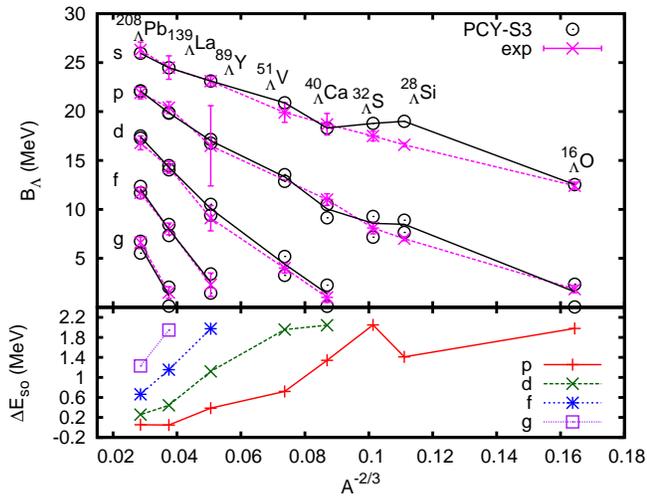}
\end{center}
\caption{(Color online) Same as Fig. \ref{fig:PCY-S1}, but with 
the parameter set PCY-S3.}
\label{fig:PCY-S3}
\end{figure}

Let us now discuss the calculated spin-orbit splittings, $\Delta E_{\rm so}$. 
These are estimated as a difference of $\Lambda$ single-particle energies 
between spin-orbit partners,  $\Delta E_{\rm so}=
\epsilon_{\Lambda,j=l-1/2}-\epsilon_{\Lambda,j=l+1/2}$,  
when the $\Lambda$ particle is put in the lowest $s$-orbital. 
Those obtained with PCY-S1 and PCY-S2 are shown in 
the lower panels of Figs. 
\ref{fig:PCY-S1} and \ref{fig:PCY-S2}, respectively. 
For both the parameter sets, although the absolute values of $\Delta E_{\rm so}$ 
are smaller by roughly a factor of 10 than those for nucleon, 
$\Delta E_{\rm so}$ alters 
its sign depending on the mass number. 
This does not happen in the meson exchange models (See Refs.\cite{SuTo,SoYaLuMe}), 
and one may consider this inversion somewhat ill-favored. 
We mention, however, that at present 
there have been no experimental data 
which exclude the possible inversion of the spin-orbit splitting 
in the medium and heavier 
mass region. 

If we exclude 
the spin-orbit splitting of the $1p$ state of $\Lambda$ in 
$^{16}_{\ \Lambda}{\rm O}$ from the fitting, that is, 
if we fit only the energy centroid of each spin-orbit partner, 
we obtain Fig. 
\ref{fig:PCY-S3} for the lambda binding energies and the spin-orbit splittings. 
The parameters for this set, PCY-S3, are summarized in 
Table \ref{tb:PCY-S3}. 
For this parameter set, 
the vector and scalar couplings 
of $\Lambda$ to nucleon remain 
natural, but the tensor coupling is 
far smaller than the expected value in Eq. (\ref{eq:2/3}). 
Since there is no constraint on the value of spin-orbit splitting, 
this parameter set yields unacceptably large 
spin-orbit splitting, 
some of them stretching even beyond the 
experimental uncertainties ({\it i.e.,} the upper bounds). 
Again, a few of them have negative values. 

Lastly, we examine the role played by the derivative terms in the Lagrangian. 
In Ref. \cite{FuSe}, it was pointed out that only one derivative term 
is well constrained by the bulk nuclear observables, {\it i.e.}, inclusion of 
a single derivative term is sufficient to obtain a good fit. 
Finelli {\it et al.} have shown that their model with only a scalar 
derivative coupling indeed reproduces well the data for normal nuclei 
\cite{FiKaVrWe06} and hypernuclei \cite{FiKaVrWe09}. 
Following Finelli {\it et al.} \cite{FiKaVrWe09,FiKaVrWe06}, 
we construct 
another parameter set PCY-S4 
by omitting the vector derivative term. 
The results are shown 
in Table \ref{tb:PCY-S4} and 
Fig. \ref{fig:PCY-S4}. 
One observes that the quality of the fit is as good as the other parameter 
sets. 
The agreement with the quark model prediction is 
also well, and the inversion of the spin-orbit partner 
is not seen for this force. 
Therefore, PCY-S4 provides an alternative parameter set to PCY-S1, 
where the main difference between the two interactions is whether 
the spin-orbit splitting is normal (PCY-S4) or inverted (PCY-S1). 

\begin{table}
\caption{The parameter set PCY-S4 obtained by setting the vector 
derivative coupling, $\delta_V^{(N\!\Lambda)}$, to be zero.  
$\chi^2_{\rm dof}$ is 0.92.}
\begin{center}
\begin{tabular}{ccc}
\hline\hline
coupling const. & value & $R$\\
\hline
$\alpha^{(N\! \Lambda)}_{S}$ & $-1.8594\times 10^{-4}\ {\rm MeV}^{-2}$ & 0.72\\
$\alpha^{(N\! \Lambda)}_{V}$ & $ 1.4981\times 10^{-4}\ {\rm MeV}^{-2}$ & 0.87\\
$\delta^{(N\! \Lambda)}_{S}$ & $-1.9958\times 10^{-10}\ {\rm MeV}^{-4}$ & --\\
$\delta^{(N\! \Lambda)}_{V}$ & $0$ & --\\
$\alpha^{(N\! \Lambda)}_{T}$ & $-5.5322\times 10^{-8}\ {\rm MeV}^{-3}$ & 0.71\\
\hline\hline
\end{tabular}
\end{center}
\label{tb:PCY-S4}
\end{table}

\begin{figure}
\begin{center}
\includegraphics[scale=.36,angle=-90]{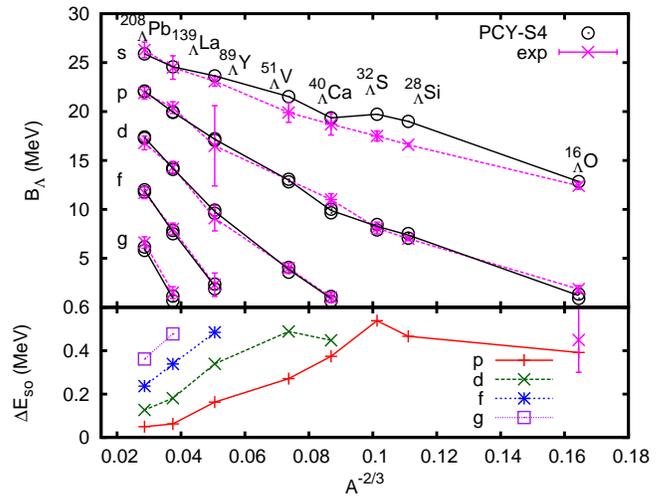}
\end{center}
\caption{(Color online) Same as Fig. \ref{fig:PCY-S1}, but with 
the parameter set PCY-S4.}
\label{fig:PCY-S4}
\end{figure}

\label{sec:result}
\end{section}

\begin{section}{SUMMARY}

We have proposed a new relativistic point coupling model 
to describe single-$\Lambda$ hypernuclei in the mean field 
approximation. 
This is a straightforward extension of the relativistic 
point coupling model for nucleons, which has a similar structure as 
the Skyrme interaction. 
To this end, we added effective contact $N$-$\Lambda$ interactions, corresponding 
to the $\Lambda$-$\sigma$ and $\Lambda$-$\omega$ couplings, to the model 
Lagrangian.  
In addition, we introduced the zero-range $N$-$\Lambda$ tensor coupling as well 
to mimic the tensor coupling between $\Lambda$ and $\omega$ meson, 
following the quark model suggestion. 

We fitted the coupling constants in the strange sector to 
the experimental data of lambda binding energies.
The four parameter sets, PCY-S1, PCY-S2, PCY-S3, and PCY-S4 were 
proposed, which well reproduce the experimental data 
through the whole mass region. 
The resulting spin-orbit splittings in PCY-S1, PCY-S2, and PCY-S4 
are smaller than that of nucleon by roughly a factor of 10 in their 
absolute values, although PCY-S3 yields too large spin-orbit splittings. 
For PCY-S1 and PCY-S2 their signs are opposite to that of nucleon 
in some nuclei in the heavier region. 
On the other hand, for PCY-S4 obtained without taking into account the 
vector derivative term, the sign of the spin-orbit splitting is the same as 
that for nucleons. 
High-precision $\gamma$-ray experiments for $\Lambda$ single-particle 
energies are awaited in order to see whether the spin-orbit splitting of heavy 
hypernuclei is normal or inverted. 

We have confirmed that 
the tensor coupling, which is ignored in the $N$-$N$ interaction, 
is quite important to reproduce the small 
spin-orbit splittings of $\Lambda$ particle. 
Without the tensor coupling, the scalar and the vector couplings 
of $\Lambda$ to nucleon are forced to be unnaturally weak  (PCY-S2). 
The tensor coupling suppresses the spin-orbit splittings, keeping the scalar 
and the vector couplings 
consistent with 
the naive quark counting. 
Those good consistency with the quark model found in 
our interaction can be a useful guide in further extending 
the point coupling model to multi-$\Lambda$ or $\Xi$ hypernuclei.

We conclude that the point coupling model is capable of 
describing single-$\Lambda$ hypernuclei as well as normal nuclei. 
The model can be an appropriate tool for relativistic calculations for 
hypernuclei on three-dimensional mesh due to its numerical simplicity. 
There has been no relativistic calculation performed on 3D mesh 
because of the variational difficulty. 
Thus the primary future work is to develop an efficient calculation 
technique for relativistic calculations on 3D mesh 
that overcomes the ``variational collapse''. 
A work in this direction is in progress. 
Further extensions of the point coupling model to multi-$\Lambda$ and $\Xi$ 
hypernuclei, and an introduction of explicit density dependences into the coupling 
constants are also interesting future works.

\label{sec:summary}
\end{section}

\begin{acknowledgments}
We thank H. Tamura, A. Ohnishi, J.A. Maruhn, and T. Koike 
for useful discussions. 
Discussions during the YIPQS international workshop 
at Yukawa Institute for Theoretical Physics, Kyoto University, 
on ``Dynamics and Correlations in Exotic Nuclei 2011 (DCEN2011)'', 
were useful to complete this work. 
This work was supported by the Global COE Program 
``Weaving Science Web beyond Particle-Matter Hierarchy'' at 
Tohoku University, 
and by the Japanese
Ministry of Education, Culture, Sports, Science and Technology
by Grant-in-Aid for Scientific Research under
the program number (C) 22540262.
\end{acknowledgments}

\begin{appendix}

\section{Center of Mass Energy}

In this Appendix, we give an explicit expression for the numerator of Eq. (\ref{cmenergy}). 
Although it has already been given in Ref. \cite{BeRuReMa}, 
we have found a few typos in their expression for 
the relativistic case with spherical symmetry. 
Here we shall correct the typos and give 
the correct formula. 

With the spherical symmetry, 
single-particle (s.p.) wave functions are given as 
\begin{equation}
\psi_{\alpha m}(\vec r)=
\left(\begin{array}{c}
\psi_{\alpha}^{(+)}(r)\mathscr{Y}_{\ell_{\alpha}^{(+)}j_{\alpha}m}(\phi,\theta)\\
i\psi_{\alpha}^{(-)}(r)\mathscr{Y}_{\ell_{\alpha}^{(-)}j_{\alpha}m}(\phi,\theta)
\end{array}\right),
%\label{eq:}
\end{equation}
where $\alpha$ is a shorthanded notation for $\{n_{\alpha},\ell^{(+)}_{\alpha},j_{\alpha}\}$, 
while $m=j_z$. Here, 
$n_{\alpha}$ is the principal quantum number, 
$\ell^{(+)}_{\alpha}$ is the orbital angular momentum of the 
upper component of the s.p. spinor, 
and $j_{\alpha}$ is the total angular momentum. 
The spherical spinor $\mathscr{Y}_{\ell jm}$ is defined by 
$\mathscr{Y}_{\ell jm}=\sum_{m'm''}
\langle\ell m' \frac{1}{2}m''|jm\rangle Y_{\ell m}\chi_{\frac{1}{2}m''}$, where 
$\chi_{\frac{1}{2}m''}$ is the spin wave function. 
The orbital angular momentum of the lower component $\ell_{\alpha}^{(-)}$ is given 
by $\ell_{\alpha}^{(-)}=2j_{\alpha}-\ell_{\alpha}^{(+)}$.
Following Ref. \cite{BeRuReMa}, we compute 
the center of mass correction in the non-relativistic 
approximation. 
The expectation value of the squared center of mass momentum $P^2_{\rm CM}$ 
with 
a mean field many-body state reads 
\begin{widetext}
\begin{equation}
\begin{array}{rcl}
\displaystyle \langle P^2_{\rm CM} \rangle&=&
\displaystyle -\hbar^2\sum_{\alpha}w_{\alpha}\sum_m
\langle\psi_{\alpha m}|\Delta|\psi_{\alpha m}\rangle\\
&&
\displaystyle -\hbar^2\sum_{\alpha,\beta}
\Bigl(w_{\alpha}w_{\beta}+\sqrt{w_{\alpha}(1-w_{\alpha})w_{\beta}(1-w_{\beta})}\Bigr)
\sum_{m,m'}|\langle\psi_{\alpha m}|\vec\nabla|\psi_{\beta m'}\rangle|^2,
\end{array}
%\label{eq:}
\end{equation}
with
\begin{equation}
\sum_m\langle\psi_{\alpha m}|\Delta|\psi_{\alpha m}\rangle
=(2j_{\alpha}+1)\sum_{\eta=\pm}\int dr\ r^2\psi_{\alpha}^{(\eta)}
\biggl[\frac{\partial^2}{\partial r^2}+\frac{2}{r}\frac{\partial}{\partial r}
-\frac{\ell_{\alpha}^{(\eta)}(\ell_{\alpha}^{(\eta)}+1)}{r^2}
\biggr]
\psi_{\alpha}^{(\eta)},
%\label{}
\end{equation}
and
\begin{eqnarray}
\begin{aligned}
&\sum_{m,m'}|\langle\psi_{\alpha m}|\vec\nabla|\psi_{\beta m'}\rangle|^2\\
&=(2j_{\alpha}+1)(2j_{\beta}+1)\sum_{\eta,\eta'}(-)^{\ell_{\alpha}^{(\eta)}+\ell_{\beta}^{(\eta')}+1}
\left\{\begin{array}{ccc}
j_{\beta} & j_{\alpha} & 1\\ 
\ell^{(\eta)}_{\alpha} & \ell_{\beta}^{(\eta)}& \frac{1}{2}
\end{array}\right\}
\left\{\begin{array}{ccc}
j_{\beta} & j_{\alpha} & 1\\ 
\ell^{(\eta')}_{\alpha} & \ell_{\beta}^{(\eta')}& \frac{1}{2}
\end{array}\right\}\\
&\times
\biggl[\delta_{\ell_{\alpha}^{(\eta)},\ell_{\beta}^{(\eta)}+1}\sqrt{\ell_{\alpha}^{(\eta)}}
\int dr\ r^2\psi^{(\eta)}_{\alpha}
\biggl(\frac{\partial}{\partial r}-\frac{\ell_{\beta}^{(\eta)}}{r}\biggr)\psi_{\beta}^{(\eta)}
%\biggr.\\
%\biggl.&
-\delta_{\ell_{\beta}^{(\eta)},\ell_{\alpha}^{(\eta)}+1}\sqrt{\ell_{\beta}^{(\eta)}}
\int dr\ r^2\psi^{(\eta)}_{\alpha}
\biggl(\frac{\partial}{\partial r}+\frac{\ell_{\beta}^{(\eta)}+1}{r}\biggr)\psi_{\beta}^{(\eta)}
\biggr]\\
&\times
\biggl[\delta_{\ell_{\beta}^{(\eta')},\ell_{\alpha}^{(\eta')}+1}\sqrt{\ell_{\beta}^{(\eta')}}
\int dr\ r^2\psi^{(\eta')}_{\beta}
\biggl(\frac{\partial}{\partial r}-\frac{\ell_{\alpha}^{(\eta')}}{r}\biggr)\psi_{\alpha}^{(\eta')}
%\biggr.\\
%\biggl.&
-\delta_{\ell_{\alpha}^{(\eta')},\ell_{\beta}^{(\eta')}+1}\sqrt{\ell_{\alpha}^{(\eta')}}
\int dr\ r^2\psi^{(\eta')}_{\beta}
\biggl(\frac{\partial}{\partial r}+\frac{\ell_{\alpha}^{(\eta')}+1}{r}\biggr)\psi_{\alpha}^{(\eta')}
\biggr],
\end{aligned}
\label{eq:exchange}
\end{eqnarray}
\end{widetext}
where $w_{\alpha}$ is the occupation probability of the level $\alpha$.  

\end{appendix}


\begin{thebibliography}{99}

\bibitem{SeWa}B. D. Serot and J. D. Walecka, 
Adv. Nucl. Phys. \textbf{16}, 1, (Plenum Press, New York, 1986).

\bibitem{VALR05}
D. Vretenar, A. Afanasjev, G.A. Lalazissis, and P. Ring, 
Phys. Rep. {\bf 409}, 101 (2005). 

\bibitem{P96}P. Ring, Prog. Part. Nucl. Phys. {\bf 37}, 
193 (1996). 

\bibitem{MTZZLG06}J. Meng, H. Toki, S.G. Zhou, 
S.Q. Zhang, W.H. Long, and L.S. Geng, 
Prog. Part. Nucl. Phys. {\bf 57}, 
470 (2006). 

\bibitem{MaMa}P. Manakos and T. Mannel, 
Z. Phys. A \textbf{330}, 223 (1988).

\bibitem{NiHoMa}B. A. Nikolaus, T. Hoch, and D. G. Madland, 
Phys. Rev. C \textbf{46}, 1757 (1992).

\bibitem{BuMaMaRe}T. B\"urvenich, D. G. Madland, J. A. Maruhn, and P.-G. Reinhard, 
Phys. Rev. C \textbf{65}, 044308 (2002).

\bibitem{MaBuMa}J. A. Maruhn, T. B\"urvenich, and D. G. Madland, 
J. Comput. Phys. \textbf{169}, 238 (2001).

\bibitem{SuBuMaReGr}A. Sulaksono, T. B\"urvenich, J. A. Maruhn, P.-G. Reinhard, and W. Greiner, 
Ann. Phys. (NY) \textbf{306}, 36 (2003).

\bibitem{NiVrRi}T. Nik\v si\'c, D. Vretenar, and P. Ring, Phys. Rev. C {\bf 78}, 034318 (2008).

\bibitem{YMRV10}J.M. Yao, J. Meng, P. Ring, and D. Vretenar, 
Phys. Rev. C {\bf 81}, 044311 (2010). 

\bibitem{NLVPMR09}
T. Nik\v si\'c, D. Vretenar, and P. Ring, Phys. Rev. C {\bf 73}, 034308 (2006);  
T. Nik\v si\'c, D. Vretenar, and P. Ring, Phys. Rev. C{ \bf 74}, 064309 (2006); 
T. Nik\v si\'c, Z.P. Li, D. Vretenar, L. Pr\'ochniak, 
J. Meng, and P. Ring, Phys. Rev. C {\bf 79}, 034303 (2009). 

\bibitem{ZhLiYaMe}P. W. Zhao, Z. P. Li, J. M. Yao, and J. Meng, 
Phys. Rev. C {\bf 82}, 054319 (2010).

\bibitem{BFHKW85}P. Bonche, H. Flocard, P.-H. Heenen, 
S.J. Krieger, and M.S. Weiss, Nucl. Phys. {\bf A443}, 39 (1985). 

\bibitem{ev8}P. Bonche, H. Flocard, and P.-H. Heenen, 
Comput. Phys. Commun. {\bf 171}, 49 (2005). 

\bibitem{ZhLiMe}Y. Zhang, H. Liang, and J. Meng, 
Int. J. Mod. Phys. E \textbf{19} 55 (2010).

\bibitem{HagiTani}K. Hagino and Y. Tanimura, 
Phys. Rev. C \textbf{82}, 057301 (2010).

\bibitem{WaKu81}H. Wallmeier and W. Kutzelnigg, Chem. Phys. Lett. \textbf{78}, 341 (1981).
\bibitem{WaKu83}H. Wallmeier and W. Kutzelnigg, Phys. Rev. A \textbf{28}, 3092 (1983).
\bibitem{StHa}R. E. Stanton and S. Havriliak, J. Chem. Phys. \textbf{81}, 1910 (1984).
\bibitem{HiKr}R. N. Hill and C. Krauthauser, Phys. Rev. Lett. \textbf{72}, 2151 (1994).
\bibitem{FaFoCh}P. Falsaperla, G. Fonte, and J. Z. Chen, Phys. Rev. A \textbf{56}, 1240 (1997).

\bibitem{MyHa08}Myaing Thi Win and K. Hagino, 
Phys. Rev. C \textbf{78}, 054311 (2008).

\bibitem{BiEnSh}Bing-Nan Lu, En-Guang Zhao, and Shan-Gui Zhou, 
Phys. Rev. C \textbf{84}, 014328 (2011). 

\bibitem{MyHaKo}Myaing Thi Win, K. Hagino, and T. Koike, 
Phys. Rev. C \textbf{83}, 014301 (2011).

\bibitem{YLHWZM11}J.M. Yao, Z.P. Li, K. Hagino, 
M. Thi Win, Y. Zhang, and J. Meng, 
Nucl. Phys. {\bf A868-869}, 12 (2011). 

\bibitem{IsKiDoOh}M. Isaka, M. Kimura, A. Dote, and A. Ohnishi, 
Phys. Rev. C \textbf{83}, 044323 (2011).

\bibitem{HiKaMiMo}E. Hiyama, M. Kamimura, K. Miyazaki, and T. Motoba, 
Phys. Rev. C \textbf{59}, 2351 (1999).

\bibitem{Ta}K. Tanida, et al., 
Phys. Rev. Lett. \textbf{86}, 1982(2001).

\bibitem{SuTo}Y. Sugahara and H. Toki, 
Prog. Theor. Phys. \textbf{92}, 803 (1994).

\bibitem{SoYaLuMe}C. Y. Song, J. M. Yao, H. F. L\" u, and J. Meng, 
Int. J. Mod. Phys. E \textbf{19}, 2538 (2010).


\bibitem{CoWe}J. Cohen and H. J. Weber, 
Phys. Rev. C \textbf{44}, 1181 (1991).

\bibitem{Jeetal}B. K. Jennings Phys. Lett. \textbf{B246}, 325 (1990); 
M. Chiapparini, A. O. Gattone, and B. K. Jennings, Nucl. Phys. \textbf{A529}, 589 (1991).

\bibitem{FiKaVrWe09}P. Finelli, N. Kaiser, D. Vretener, and W. Weise, 
Nucl. Phys. \textbf{A831}, 163 (2009); 
P. Finelli, Nucl. Phys. \textbf{A835}, 418 (2010).

\bibitem{BrWe}R. Brockmann and W. Weise, 
Phys. Lett. \textbf{69B}, 167 (1977).

\bibitem{BoBo}J. Boguta and S. Bohrmann, 
Phys. Lett. \textbf{102B}, 93 (1981).

\bibitem{Bou}A. Bouyssy, Nucl. Phys. \textbf{A381}, 445 (1982).

\bibitem{MaJe}J. Mare\v s and B. K. Jennings, 
Phys. Rev. C \textbf{49}, 2472 (1994).

\bibitem{V98}D. Vretenar, W. P\"oschl, G.A. Lalazissis, 
and P. Ring, Phys. Rev. C{\bf 57}, R1060 (1998). 

\bibitem{TsMaMaOh}K. Tsubakihara, H. Maekawa,  H. Matsumiya, and A. Ohnishi, 
Phys. Rev. C \textbf{81}, 065206 (2010).

\bibitem{HaTa}O. Hashimoto and H. Tamura, 
Prog. Part. Nucl. Phys. \textbf{57}, 564 (2006).

\bibitem{Mo}T. Motoba, Nucl. Phys. \textbf{A639}, 135c (1998).

\bibitem{AlFeRo}B. Alder, S. Fernbach, and M. Rotenberg, 
\textit{Methods in Computational Physics vol. 6 Nuclear Physics}
 (Academic Press, New York and London, 1966).
 
\bibitem{UsBo}Q. N. Usmani and A. R. Bodmer
Phys. Rev. C \textbf{60}, 055215 (1999).

\bibitem{FuSe}R. J. Furnstahl and Brian D. Serot, Nucl. Phys. {\bf A671}, 447 (2000).

\bibitem{FiKaVrWe06}P. Finelli, N. Kaiser, D. Vretener, and W. Weise, 
Nucl. Phys. \textbf{A770}, 1 (2006).

\bibitem{BeRuReMa}M. Bender, K. Rutz, P.-G. Reinhard, and J.A. Maruhn
, Eur. Phys. J. A {\bf 7}, 467 (2000).

\end{thebibliography}
\end{document}